\documentclass[11pt,preprint]{aastex}

\usepackage{graphicx}

\newcommand\ergsec{{~erg~s$^{-1}$}}
\newcommand\ergcmsa{{~erg~cm$^{-2}$~s$^{-1}$~\AA$^{-1}$}}
\newcommand\kms{km~s$^{-1}$}
\newcommand\ghz{\rm GHz}
\newcommand\mhz{\rm MHz}
\newcommand\Whz{\rm W~Hz$^{-1}$}
\newcommand\eg{{\it e.g.~\/}}
\newcommand\ha{{H$\alpha$}}
\newcommand\hb{{H$\beta$}}
\newcommand\feii{[Fe\,{\small~II}]}
\newcommand\oiii{[O\,{\small~III}]}
\newcommand\pfeii{Fe\,{\small~II}}
\newcommand\pcaii{Ca\,{\small~II}}
\newcommand\fullmgi{[Mg\,{\small~I]}~$\lambda$5175} 
\newcommand\fulldnii{[N\,{\small~II}]$\,\lambda\lambda$6548,6583}

\newcommand\fulloiii{[O\,{\small~III}]$\,\lambda$5007}
\newcommand\fullpfei{Fe\,{\small~I}~$\,\lambda$5269}
\newcommand\fullnii{[N\,{\small~II}]$\,\lambda$6583}
\newcommand\nii{[N\,{\small~II}]}
\newcommand\fullsii{[S\,{\small~II}]$\,\lambda\lambda$6716,6731}
\newcommand\fulloi{[O\,{\small~I}]$\,\lambda$6300}

\slugcomment{Accepted for publication in The Astronomical Journal}

\begin{document}

\title{Exploring NLS1 through the physical properties of their hosts\footnote{
Partially based
on observations made with the Asiago 1.82 m telescope of the Padova 
Astronomical Observatory}}
\author{V. Botte\altaffilmark{2}, S. Ciroi\altaffilmark{2}, P. Rafanelli
and F. Di Mille}
\affil{\it Department of Astronomy, University of Padova, 
vicolo dell'Osservatorio 2, I-35122 Padova, Italy}
\altaffiltext{2}{Guest investigator of the UK Astronomy Data Centre}

\begin{abstract}

In this work we aim at addressing the still open question about the
nature of Narrow-Line Seyfert 1 (NLS1) galaxies: are they really active nuclei
with lower mass Black-Holes (BHs) than Seyfert 1 (S1) and quasars?
Our approach is based on the recently discovered physical connections
between nuclear supermassive BHs and their hosting spheroids (spiral bulges or
ellipticals). In particular we compare BH masses of NLS1s and S1s analyzing the 
properties of their hosts by means of spectroscopic and photometric data in the 
optical wavelength domain.
We find that NLS1s fill the low BH mass and bulge luminosity values of the 
$\rm {\cal M}_{BH}-M_{B}$ relation, a result strongly suggesting that NLS1s are 
active nuclei where less massive BHs are hosted by less massive bulges. 
The correlation is good with a relatively small scatter fitting simultaneously 
NLS1s, S1s and quasars.
On the other hand, NLS1s seem to share the same stellar velocity dispersion 
range of S1s in the $\rm {\cal M}_{BH}-\sigma_*$ relation, indicating that 
NLS1s have a smaller BH/bulge mass ratio than S1s. 
These two conflicting results support in any case the idea that NLS1s could be
young S1s.  
Finally we do not confirm the significantly non linear BH--bulge relation 
claimed by some authors.

\end{abstract}



\section{INTRODUCTION}

Narrow-Line Seyfert 1 (NLS1) galaxies belong to the family of active galactic 
nuclei (AGNs) and owe their name to the peculiar properties that distinguish 
them from the other type 1 AGNs.
In general a Seyfert 1 (S1) is classified as NLS1 when its nuclear spectrum 
shows the permitted lines only slightly broader (FWHM(H$\beta) < 2000$ km 
s$^{-1}$) than the forbidden ones \citep{op85}.
But NLS1 galaxies are characterized also by a strong Fe II emission, a
soft X-ray slope steeper than the slope typical for S1 galaxies,
rapid and large soft X-ray variability \citep*{bolet96},
a weak Big Blue Bump in the optical/UV range, likely shifted toward higher
energies \citep{pou87}, a bright IR emission and
a nuclear super-solar metallicity \citep{mathur00,km01}.

Even though NLS1s are known since almost twenty years \citep{op85}, their 
nature is still a matter of debate. 
For example, at present the reason of the narrowness of broad permitted lines 
in NLS1s is not clear. 
\citet{bolet96} suggested that if the gravitational force from the 
central Black-Hole (BH) is the main cause of motion of the Broad Line Region 
(BLR) clouds, narrower optical emission lines will result from smaller 
$\cal M_{\rm BH}$ ($\sim 10^6-10^7 {\cal M}_{\odot}$) provided 
the BLR distance from the central source does not change 
strongly with $\cal M_{\rm BH}$. These BHs with smaller masses are
expected to accrete matter at near or super-Eddington rates in order to
maintain the relatively normal observed luminosities.
Recently \citet{matet01} proposed that NLS1s may be relatively young AGNs 
hosting BHs still in a growing phase.
Nevertheless other authors suggested that if the BLR clouds were largely 
confined to a plane, as it seems to happen in radio-loud AGNs   
\citep[see e.g.][and references therein]{mcdun02}, 
NLS1 galaxies could be simply a case in which their BLR is observed more 
face-on than in S1s \citep[see e.g.][]{op85}. 
\citet{smi02} pointed out that also a partly obscured BLR could justify 
narrow permitted lines, but these last two scenarios should produce 
polarization. They claimed that \ha\, polarization properties of the NLS1s 
are indistinguishable from those of S1s, excluding obscuration of the 
inner regions of BLR or a face-on orientation of a disk-like BLR as the 
explanation for the relatively narrow broad-line profiles.

The connection between physical properties of nuclear supermassive BHs  
and their host galaxies, on which several works have focused in the last years, 
might turn out to be a powerful tool to understand the nature of NLS1s and 
settle the above cited controversies.
\citet{koric95} first found out a correlation between BH mass 
($\cal M_{\rm BH}$) and the absolute B magnitude of the spheroidal component 
($M_{B}$). 
\citet{maget98} determined  $\cal M_{\rm BH}$ values for a sample of 32 nearby 
galaxies and suggested that $\cal M_{\rm BH}$ is proportional to the 
$\cal M_{\rm bulge}$ such that on average 
$\cal M_{\rm BH}/\cal M_{\rm bulge}\sim$ 0.005. Other authors estimated a 
similar mass ratio, $\sim$ 0.002 (see e.g.,  Ho 1999).
Recently, studying a sample of nearby galaxies \citet{gebet00a},
\citet{fermer00} and \citet{tre02}  have shown that $\cal M_{\rm BH}$ is tightly 
correlated with the velocity dispersion of the bulge stellar component 
($\sigma_*$), although they disagreed about the value of the slope.

In AGNs as well, BHs are also expected to correlate with 
their host bulges. This possibility was explored on a sample of PG quasars 
by \citet{lao98} who found agreement with the relation of \citet{maget98}. 
Later \citet{wan99} claimed that Seyfert galaxies show 
on average a $\cal M_{\rm BH}/\cal M_{\rm bulge}$ ratio systematically lower 
than that for normal galaxies and quasars.
Conversely \citet{gebet00b} included in their work seven AGNs for which the
$\cal M_{\rm BH}$ were obtained by means of the reverberation mapping 
technique. They found that these objects were in agreement with their 
previously found $\cal M_{\rm BH}$--$\sigma_*$ correlation. 
Further support came from \citet{mcdun01}, and from \citet{Wuhan01}
who studied samples of quasars and Seyfert galaxies, and did not find 
any evidence that Seyfert galaxies follow a different $\cal M_{\rm BH}-
\cal M_{\rm bulge}$ relation from quasars or nearby galaxies.\\
Until now no agreement has been found about NLS1s. 
\citet*{matet01} and very recently \citet{bz03} and \citet{gm04} showed that 
the $\cal M_{\rm BH}/\cal M_{\rm bulge}$ 
ratio in NLS1s is significantly smaller than that for Seyfert galaxies. 
Conversely \citet{walu01}, studying a sample of 59 NLS1s observed 
spectroscopically by \citet*{vvg01}, found that there is non clear 
difference in the $\cal M_{\rm BH}$--$\sigma_*$ relation (where $\sigma_*$ is 
represented by the \oiii\ emission line width) between NLS1s, 
Broad-Line AGNs and nearby galaxies.

Our purpose in this work is to investigate the nature of NLS1s by exploring  
the physical properties of their bulges, namely the luminosity and the nuclear 
stellar velocity dispersion.
Our approach makes use of both new observational data and data from the 
literature. The necessary corrections are applied to the latter in order to 
obtain as homogeneous as possible a dataset. Furthermore, as a main difference 
with respect to several other works on this topic, when determining the bulge 
properties of the host galaxies we take into account the influence of the 
AGN, which, as we show, can be non-negligible and affect the results 
considerably. The structure of the paper is as follows:
in \S\,2 we present our dataset and derive and compare BH mass values for a 
sample of NLS1s and S1s. In \S\,3 we estimate the blue absolute magnitudes and 
the stellar velocity dispersions of their bulges.  
Our results are summarized and discussed in \S\,4.

\section{BLACK-HOLE PROPERTIES}

\subsection{Spectroscopic data}

Firstly, we have isolated a list of 23 NLS1 and 23 S1 galaxies of the northern 
emisphere from \citet{vvg01} on the basis of their ``S1n'' and ``S1.0'' 
classification and of the redshift, z $<0.1$,  chosen to avoid that \hb\ and 
\oiii\ lines fall in a spectral region with strong night-sky emission lines. \\
No other selection criteria were applied. This sample is complete 
up to visual magnitude 15.5, corresponding to the 80 per cent of the selected 
galaxies, and therefore it is useful for our purposes.\\
Out of this sample, we were able to collect optical spectra for 22 NLS1s and 
15 S1s. 
In particular 19 NLS1s and 7 S1s were extracted from the public data 
available in the Isaac Newton Group (ING) Archive. 
These spectra were obtained for 
different purposes in 1995, 1996, 1999 and 2000, mostly with the Intermediate
Dispersion Spectrograph (IDS)  mounted at 
the 2.5m Isaac Newton Telescope (INT, Canary Islands, Spain), and the others 
with the ISIS Double Beam Spectrograph (ISIS) at the 4.2m William Herschel 
Telescope (WHT, Canary Islands, Spain).
Other 3 NLS1s and 8 S1s were observed directly by us in 2002 September and 
in 2003 January using the Asiago Faint Object Spectrograph and Camera (AFOSC) 
mounted at the 1.82m telescope of the Padova 
Astronomical Observatory (Asiago, Italy).
Tables ~\ref{tab1} and ~\ref{tab2} summarize the instrumental setup 
and the total wavelength coverage for each observation. 

All spectra were reduced with the same procedure.
The usual data reduction steps - bias and flat field corrections, 
cosmic rays removal, wavelength linearization, sky-background subtraction and 
flux calibration-  were carried out with IRAF packages\footnote{
IRAF is written and supported by NOAO (Tucson, Arizona), which is operated 
by AURA, Inc. under cooperative agreement with the National Science 
Foundation}.
A one-dimensional spectrum of the nucleus was obtained for each galaxy
summing a number of pixels along the spatial direction on the basis of the 
seeing conditions. When available, adjacent spectral ranges of the same source 
were combined together. 
Then a correction for Galactic extinction was applied using for each galaxy 
the value given by NED\footnote{NASA Extragalactic Database} and the 
\citet{caet89} extinction law.
The so-processed spectra were shifted to the rest frame and  
the intrinsic absorption was removed following the technique used by 
\citet{creet02}. In particular 
we have determined the value of the internal reddening using the relation: 
$\rm E(B-V) = 2.5~[log(X_B) - log(X_V)]$, where X is the ratio of the 
Mrk 478 continuum fit to that of the other galaxies of the sample evaluated at 
the effective wavelengths of B (4400 \AA) and V (5500 \AA) photometric bands.
Contrary to \citet{creet02} we could not use as reference Mrk 493, whose
spectrum was available only for wavelengths $>$ 4400\AA\, thus yielding 
a rather uncertain estimate of X$\rm_B$. The spectrum of Mrk 478 was used, 
instead, after correction for internal absorption by means of the hydrogen 
column density $\rm N_H=2.0\pm10^{20}~cm^{-2}$, given by \citet{bolet96}.\\
One of the peculiar features in the optical spectrum of most NLS1 galaxies 
is the presence of strong emission of \pfeii\ multiplets centered at 
4570 \AA, 5190 \AA\ and 5300 \AA .
In order to remove them and allow a more precise measure of \hb\ and \oiii\ 
emission lines we have produced a \pfeii\ template 
using the spectrum of the NLS1 I Zw 1 as suggested by \citet{boet92}. 
The template was scaled in intensity and conveniently smoothed to match 
the spectrum of each galaxy showing evident Fe II multiplets, and then 
subtracted.

\subsection{Wrong classifications}

\begin{figure}
\plotone{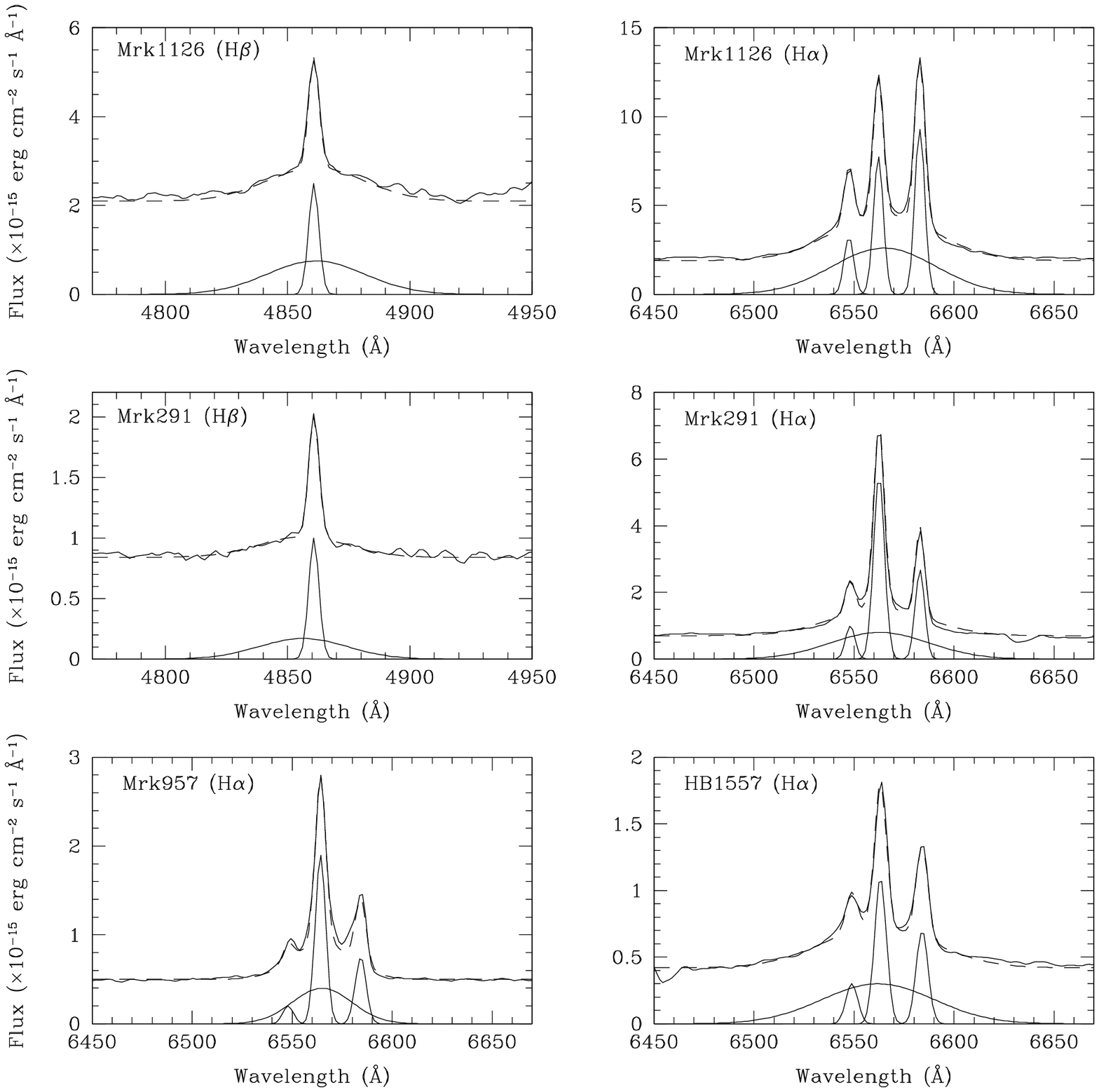}
\caption[Botte.fig1.ps]{Multigaussian deblending of broad and narrow \hb\ and \ha\ 
emission lines for the wrongly classified NLS1s. In each panel the observed profile 
(top) is compared with the fit (dashed line). Single gaussian components 
are also shown (bottom). \label{f1}}
\end{figure}

A fast inspection of each galaxy spectrum allowed us to isolate four NLS1s, 
Mrk 1126, Mrk 291, Mrk 957 and HB 1557, which in our opinion are wrongly 
classified.
After a multi-gaussian fit of \hb, and \ha+\fulldnii\ emission lines 
we can assert that these active nuclei have permitted lines with a clear 
composite broad$+$narrow profile, where the narrow components have widths 
similar to those of the forbidden lines ($\sim 250-300$ \kms) (Fig.~\ref{f1}).
This kind of profile is typically shown by intermediate Seyfert galaxies.
Indeed \citet{ost89} introduced the notation Seyfert 1.5, 1.8, 
and 1.9, to indicate the simultaneous presence of narrow and broad permitted 
lines in many spectra of Seyfert galaxies.

Broad \hb\ and \ha\ components were fitted in Mrk 1126 and Mrk 291, obtaining 
widths of about 2600-2900 \kms\ , while  only  \ha-broad was detected in Mrk 957 
and HB 1557, and fitted to widths of about 1600 and 3000 \kms\
respectively. We observe that the value of  \ha-broad in Mrk 957 is lower than 
the others. This could be caused by a non reliable multigaussian fit of
\ha+\nii\ blend. 
Indeed we have the impression that a blue \nii\ component exists in
addition to the others already identified, but the spectral resolution is not 
sufficiently high to obtain a stable fit of the profile with six or more 
gaussians. Moreover we noted that Mrk 957 is the only out of the four galaxies 
showing Fe II multiplets, whose strong emission is one of the typical features 
of NLS1s.

Therefore we excluded these four galaxies from our sample of NLS1s, but we
included UGC 3478 until now wrongly classified as S1. UGC 3478 
has an optical nuclear spectrum typical of NLS1s: we have found narrow Balmer 
emission lines (FWHM(\hb)=1600 \kms), a low \fulloiii/\hb\ ratio (=4.8), and a 
strong emission of \feii\ multiplets. Moreover the luminosities of the low 
ionization emission lines, \fullnii, \fullsii\ and \fulloi, assume weak values.
The ASCA hard X-ray spectrum, available in TARTARUS database, shows a steep 
power law distribution with photon index $\Gamma_X\sim2.3$ similar to that 
given by \citet{Lei99} for NLS1s ($\Gamma_X\sim2.19\pm0.10$).

\subsection{BH masses estimation}

\begin{figure}
\plotone{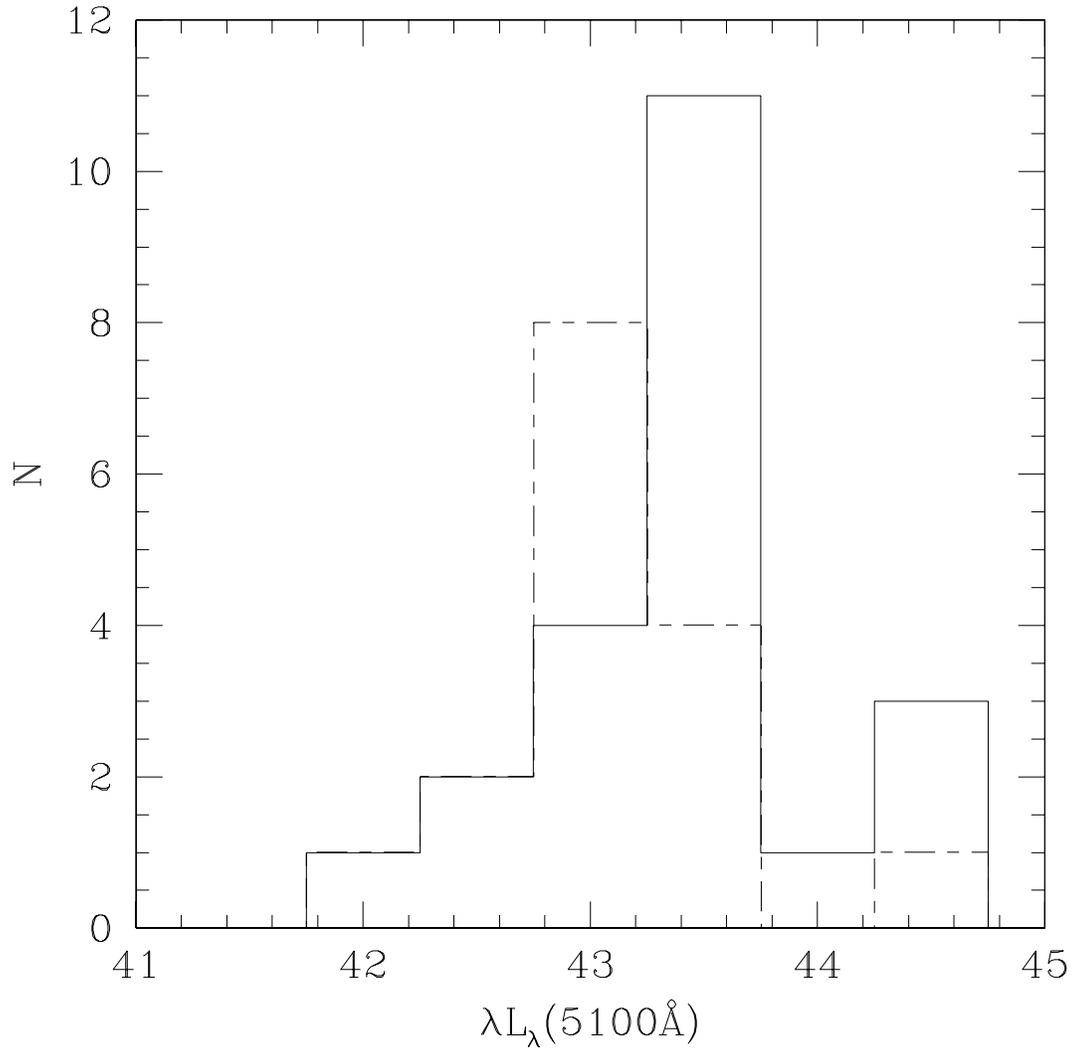}
\caption[Botte.fig2.ps]{Comparison between the distributions of the measured nuclear 
continuum luminosity for NLS1 (solid) and S1 (dashed) galaxies. 
The median values are 43.51$\pm$0.65 and 43.08$\pm$0.45 respectively. \label{f2}}
\end{figure}

The stellar dynamical techniques to derive BH masses are severely
limited for AGNs, since they require high S/N measurements of stellar
absorption features which are often lost in the glare of a bright active
nucleus \citep{ne00}.
The alternative solution consists in applying the virial theorem to the BLR 
clouds, gravitationally bound to the central mass and
located at distances of few light-days (S1s) to several light-weeks (quasars): 

\begin{equation}
\rm {\cal M}_{BH}={R_{BLR}~V^2~G^{-1}}
\label{BLR}
\end{equation}

where $\rm R_{BLR}$ is the radius of the BLR, V the velocity of the broad-line
emitting gas, and G the gravitational constant.
Even if there is no general consensus about the dynamics of the
BLR, evidence for Keplerian motions of the BLR clouds was found by
\citet{petwan99} and \citet{wanet99} using the reverberation mapping technique.
This is one of the major tools for studying correlated 
variations of the lines and continuum emission of AGNs and determine 
the size ($\rm R_{BLR}$) and the geometry of the BLR \citep[see
e.g.,][]{pet93}.\\
However, the lack of long term variability monitoring makes it difficult to 
measure $\rm R_{BLR}$ of most Seyfert galaxies using this method.
As an alternative, $\rm R_{BLR}$ can be estimated by the empirical 
relationship between the BLR size and the luminosity of the continuum at 
5100\AA\ found by \citet{kaspi00}:

\begin{equation}
\rm R_{BLR}=32.9_{-1.9}^{+2.0}\biggl[\frac{\lambda L_{\lambda}(5100\AA)}
{10^{44}~erg~s^{-1}}\biggr]^{0.700\pm0.033}
\label{kr}
\end{equation}
Since the BLR consists of photoionized clouds of gas and the luminosities of 
NLS1s and S1s are generally comparable,(see e.g. \citealt{padraf88}; 
\citealt{bolet96}), we applied this relation, whose validity for a few NLS1s was 
proved by \citet{pet00}.
Indeed after having measured $\lambda \rm L_{\lambda}$
(5100\AA)\footnote{Throughout this paper we assume $\rm H_0=75~km ~
s^{-1}~Mpc^{-1}$.} for each object of our sample we have verified that NLS1s and S1s 
have similar optical luminosity of continuum (log $\lambda \rm L_{\lambda}$
(5100\AA) = 43.51$\pm$0.65 vs. 43.08$\pm$0.45 respectively), as can be seen 
in Fig.~\ref{f2}.\\
Some authors produced substantial evidence that the broad-line emitting 
material has a flattened disk-like geometry in radio-loud quasars 
\citep[see e.g., ][]{veset00}. 
Such evidence is less strong for radio-quiet AGNs, even if very recently 
\citet{stra03} found that a high percentage of AGNs showing double-peaked 
Balmer lines are radio-quiet.
Therefore, we have estimated 
the parameter V from the emission-line width of \hb\ by assuming that the 
velocity dispersion in the line emitting gas is isotropic and after having 
removed the instrumental width:

\begin{equation}
\rm V=({\sqrt3}/{2})FWHM(H\beta)
\end{equation}

The resulting $\cal M_{\rm BH}$ values are listed in Tables ~\ref{tab6} 
and \ref{tab7}. As expected NLS1s have on average BHs with smaller masses than 
S1s: log $\cal M_{\rm BH}$ = 6.65$\pm$0.64 vs. 7.37$\pm$0.62, a result which 
depends directly on the narrowness of the Balmer emission lines, since
nuclear luminosities and therefore BLR radii are quite similar in 
both samples. Similar considerations can be found in a contemporary paper by 
\citet{gm04}.\\
Of course such $\cal M_{\rm BH}$ values are characterized by 
some uncertainties, and we have tried to analyze them.
First of all the uncertainties given in equation (\ref{kr}) produce errors 
$< 15$ per cent concerning $\rm R_{BLR}$ calculated values. 
Second, the luminosity of 
the continuum is affected by flux calibration errors and by intrinsic 
varibility of the sources. About this last point, following the discussion 
by \citet{walu01}, who noticed that the continuum variation is not larger 
than a factor of two for most AGNs, we assumed a $30$ per cent error 
as upper limit. The spectrophotometry accuracy, evaluated through multiple
observations of standard stars, was estimated around $10-20$ per cent for 
archival data, and $20-30$ per cent for Asiago data.
Combining these errors together we obtain a $< 40$ per cent error for 
$\rm R_{BLR}$.
The final uncertainty of $\cal M_{\rm BH}$ is likely less or around 
$50$ per cent, corresponding to $\sim 0.25$ dex in logarithmic scale.
It should be taken into account that a non negligible fraction of optical 
light may come from the host galaxy, especially in case of low luminosity AGNs. 
Since a precise estimate of the host contribution is not straightforward, we 
decided to neglect it and remain closer to the procedure followed by 
\citet{kaspi00} to obtain their empirical relation.

\section{BULGE PROPERTIES}

After having calculated the masses of the BHs hosted by each galaxy of 
our samples of NLS1s and S1s, we have investigated the physical properties of
their bulges deriving their blue luminosities and the values of the central
stellar velocity dispersion.
Then we have explored the connection between bulges and nuclear BHs in order to
find out where NLS1s are placed with respect to S1s.

\subsection{Blue luminosity}

Since we did not have at our disposal photometric data to measure B-band 
magnitudes ($m_{\rm B}$), we took from literature, when available, the values
for the objects of our samples 
\citep{win97,mac90,graet93,schiki00,pruhe98}.
When we found multiple estimates we calculated a median value.\\

The $m_{\rm B}$ were corrected for Galactic extinction ($\Delta{m_G}$), taking 
values given by \citet{sch98}, and internal absorption ($\Delta{m_i}$), 
following the relation given in the introduction to the Third Reference 
Catalogue of Bright Galaxies \citep[][ RC3]{deV91} : $\Delta m_i = \alpha_{\rm
T} ~log(\sec~i)$, where $\rm \alpha_T = 1.5 - 0.03~(T-5)^2 $ for spiral galaxies 
(T$\geq$0). Inclination values $i$ were extracted from 
HyperLeda\footnote{http://leda.univ-lyon1.fr/} database.
The K correction was also applied ($\Delta{m_k}$), again by following 
the method described in RC3.

Contrary to what done by several other authors, we decided to take 
into account the emission-line and non-stellar continuum contributions from the 
AGN to the total magnitude of each galaxy. 
To do this we followed the approximations given by \citet{whi92}. In particular 
we used the formulae (2), (3), (4) and (5) given in that paper for $\rm F_{cF}$ 
and $\rm F_{cH}$, which are the effective continuum fluxes in B-band due to the 
forbidden and Balmer emission lines.
Instead, before calculating the nonstellar continuum flux $\rm F_{cC}$, we 
noticed that \citet{whi92} assumed a typical value of $\sim$100\AA\ as  
equivalent width of \hb\ . Since we have the spectra and the underline 
continuum of \hb\ is very similar to 
the already measured continuum at 5100\AA, we could take as approximation 
$\rm F_{cC}\simeq F_\lambda(5100\AA)$, having adopted a power law with spectral 
index $\alpha=-1.0$ as done by \citet{whi92}.\\
The total non stellar flux $\rm F_c=F_{cF}+F_{cH}+F_{cC}$ gives the required 
correction $\Delta{m_A}$ to be applied to $m_{\rm B}$. 
The final corrected total magnitudes are given by:
$m_B^{\prime}=m_B+\Delta{m_A}-\Delta{m_i}-\Delta{m_G}-\Delta{m_k}$.

\begin{figure}
\plotone{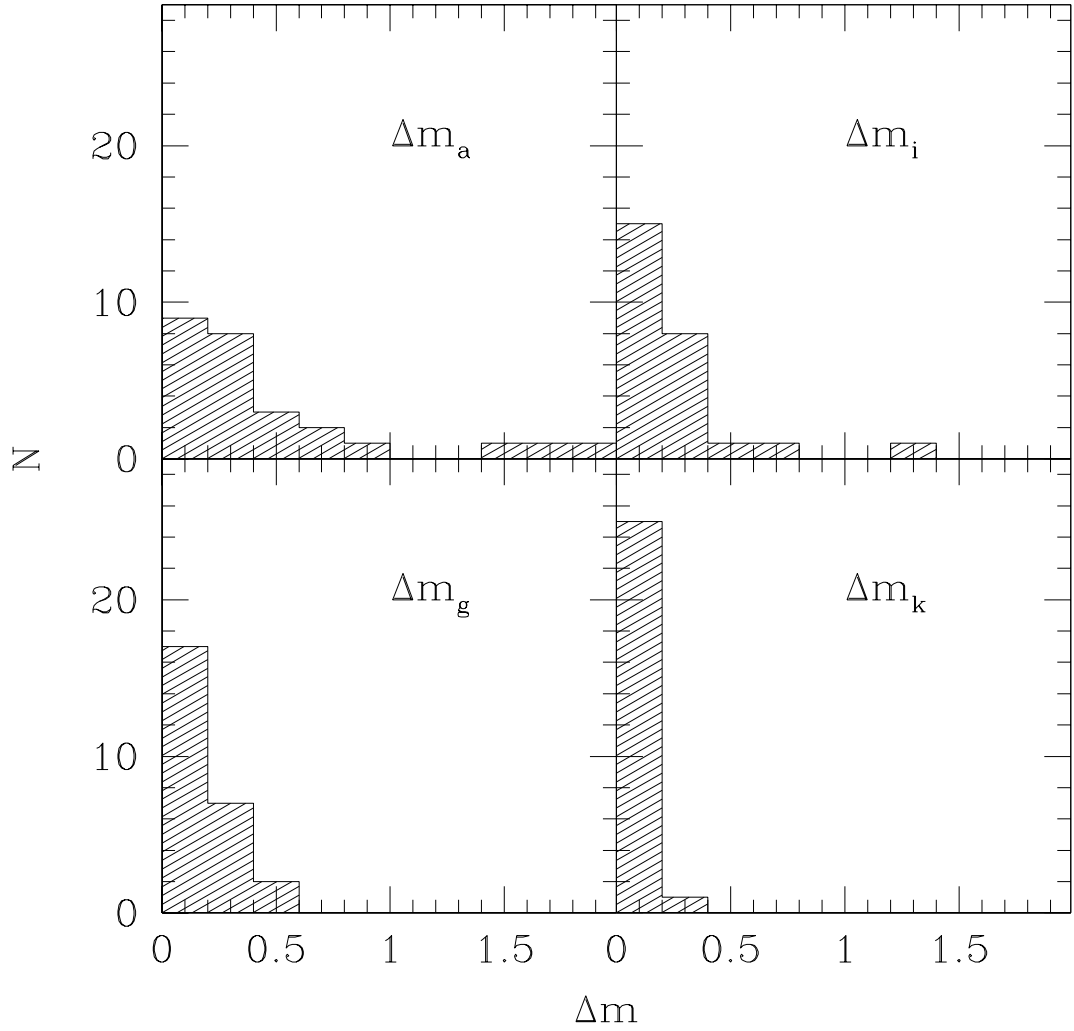}
\caption[Botte.fig3.ps]{Distribution of magnitude correction terms applied to $m_{\rm B}$
values. \label{f3}}
\end{figure}

The distribution of these terms are shown in Fig.~\ref{f3}. 
It can be easily noticed that $\Delta{m_k}$ plays a minor role in the total 
contribution to the magnitude corrections, while $\Delta{m_A}$ is as 
important as ---and sometimes even more important--- $\Delta{m_i}$ and 
$\Delta{m_g}$. This strongly indicates that neglecting $\Delta{m_A}$ can be a
dangerous approximation when bright AGNs are considered.
 
After having converted $m_{\rm B}^{\prime}$ into absolute magnitudes 
$M_{\rm B}$, listed in Table~\ref{tab3}, we obtained the bulge magnitudes 
($M_{\rm B,bulge}=M_{\rm B}+\Delta m_{bulge}$) by applying the formula given 
by \citet{side86}, who found a 
relation between the bulge-to-total (B/T) luminosity ratio and the 
morphological type: $\Delta m_{bulge} = \rm
0.324~(T+5)-0.054~(T+5)^2+0.0047~(T+5)^3$.

The morphology of each galaxy was extracted from NED and checked by
visual inspection of the POSS II digitized images, for lower redshift sources, 
which did not require high spatial resolution,  and of the HST public images, 
when available in the archive, for higher redshift galaxies.
We excluded the sources for which a "compact" classification was given.\\
The resulting magnitudes show that NLS1s have typically lower luminosity 
bulges than S1s: \\
$M_{\rm B,bulge}(\rm NLS1) = -18.54\pm1.06$ vs. 
$M_{\rm B,bulge}(\rm S1) = -19.80\pm0.73$.

Giving a realistic estimate of the magnitude errors is not an easy task.
All our photometric data are taken from literature, based on CCD 
observations, and given mostly with accuracy $< 0.05$ mag.
But our $\rm M_{B}$ values are affected by additional errors, which are 
introduced by the application of the correction terms: $\Delta m_A,~\Delta m_i,
~\Delta m_G,~\Delta m_k$ and $\Delta m_{bulge}$.\\
Among them, $\Delta m_i$ and $\Delta m_k$ are both dependent on the 
morphological type T, and the error caused by a wrong classification will be 
typically lower than 0.1 mag.
$\Delta m_{bulge}$ is also a function of T, and increases strongly for 
late-type spirals. A maximum uncertainty $\Delta T < 2$ in the 
morphological classification of our objects translates into errors $< 0.5$ 
and $< 1$ mag for early- and late-type spirals, respectively.
$\Delta m_{A}$ is dominated by the contribution of the AGN continuum, since 
the emission lines affect only $\sim$10 per cent of the total correction. 
Therefore $\Delta m_{A}$ depends on the spectrophotometric 
calibration of the continuum, and e.g. an accuracy of 20 per cent  will correspond 
to an uncertainty of $\sim$ 0.2 mag.

\begin{figure}
\plotone{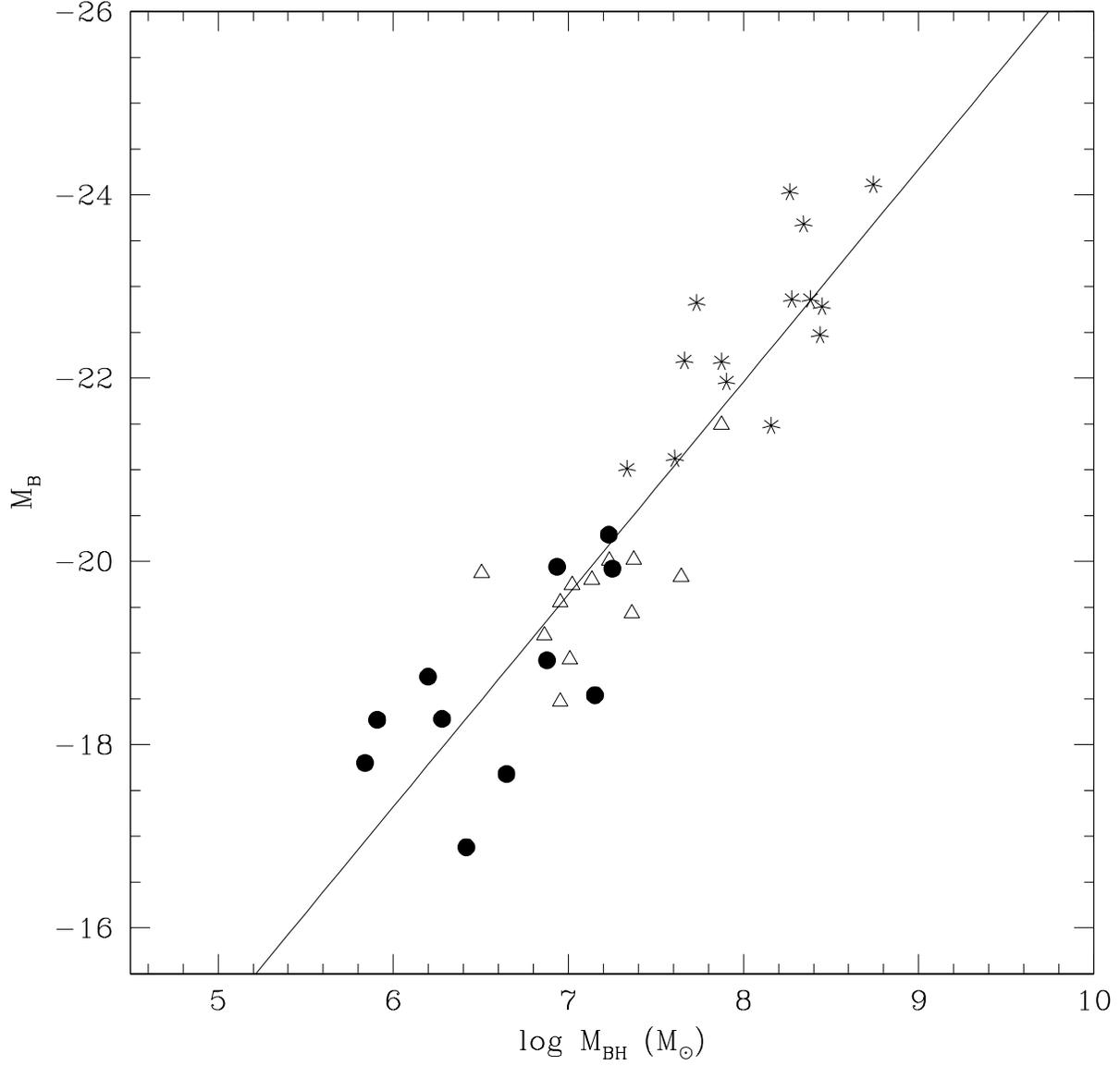}
\caption[Botte.fig4.ps]{${\cal M}_{\rm BH}-\rm M_{B}$ relation. NLS1s are
indicated by filled circles, S1s by empty triangles and quasars by asterisks.
The solid line represents a least square fit of the total sample of objects:
$\rm M_{B}~=-2.32(\pm0.18)~log({\cal M}_{BH})-3.40(\pm1.36)$. The
correlation coefficient is R = 0.91. \label{f4}}
\end{figure}

In Fig.~\ref{f4} we plotted $M_{\rm B,bulge}$ values against 
${\cal M}_{\rm BH}$. NLS1s are represented by filled dots, while S1s by 
empty triangles.
For completeness, we decided to include in this analysis a sample of 14 
quasars (asterisks), whose ${\cal M}_{BH}$ and $M_{B}$ are given 
by \citet{kaspi00}. 
The sources were selected from their list excluding those having a
${\cal M}_{BH}$ error greater than the value itself.
The absolute magnitudes were corrected for AGN contribution using only the 
measured continuum flux at 5100\AA, available in the same paper 
(Table~\ref{tab4}).
As expected after having calculated the median values of the plotted physical
quantities, NLS1s fall in the lower ranges of the plot, and are well 
separated by S1s both in BH mass and bulge luminosity values, even if a region
of overlap inevitably exists. Moreover, NLS1s, S1s and quasars seem to be well 
correlated in the plane ${\cal M}_{BH}-M_{B}$. 
Therefore we attempted a least square fit (represented by the solid line in 
Fig.~\ref{f4}) considering all points together, and obtaining: 

\begin{equation}
\rm M_{B}~=-2.32(\pm0.18)~log({\cal M}_{BH})-3.40(\pm1.36)
\label{magni}
\end{equation}

with a correlation coefficient R$=0.91$.

In order to check the consistence of this result, we first converted blue absolute 
magnitudes into V band, by applying the morphological type dependent B--V values 
given by \citet[][their Table 4]{jw03}, and we did a new fit: 
$\rm M_{V}~=-2.48(\pm0.19)~log({\cal M}_{BH})-3.09(\pm1.40)$.
Then we applied the standard relation: $\rm M_{V}~=+4.83 -
~2.5~log(L_V/L_{\odot})$ and a mass-to-luminosity conversion for bulges 
and spheroidal galaxies found by \citet{maget98}: $\rm log({\cal M}/{\cal M}_\odot) = -1.11 +
1.18(\pm0.03)~log(L/L_\odot)$, obtaining as a result:

\begin{equation}
\rm log({\cal M}_{\rm BH}/{\cal M}_{\odot})=0.85(\pm0.09)~log({\cal M}_{\rm
bulge}/{\cal M}_{\odot})-2.25(\pm0.88)
\end{equation}

In Table~\ref{tab5} we compare our result with similar 
${\cal M}_{\rm BH}\propto{\cal M}_{\rm bulge}^{\alpha}$ relations investigated 
by other authors. We notice that \citet{wan02} and \citet{mcdun02} obtained 
slopes of 0.74 and 0.88 respectively, which are consistent with our
0.85$\pm$0.09 value.\\
On the contrary \citet{bz03} found a steeper relation ($\alpha = 1.61\pm0.59$)
for a sample of 22 Narrow Line AGNs. The lower number of points and the limited 
range of the bulge mass values in comparison with their scatter could be at 
the origin of this result. Indeed the slope of their relation shows an error 
significantly larger than those presented by other authors. A more stable and 
less uncertain fit can be obtained by considering also broad-line AGNs, 
therefore spanning a wider range in both the physical quantities, BH and bulge 
masses. 
\citet{lao01} also found a steeper relation (1.36$\pm$0.15).
A possible explanation could be that the \citet{kaspi00} relation is defined 
by using monochromatic luminosity, while \citet{lao98,lao01} started from 
bolometric luminosity and applied a constant to convert into 
$\rm L_{\lambda}(5100\AA)$.

\subsection{Stellar velocity dispersion}

\begin{figure}
\plotone{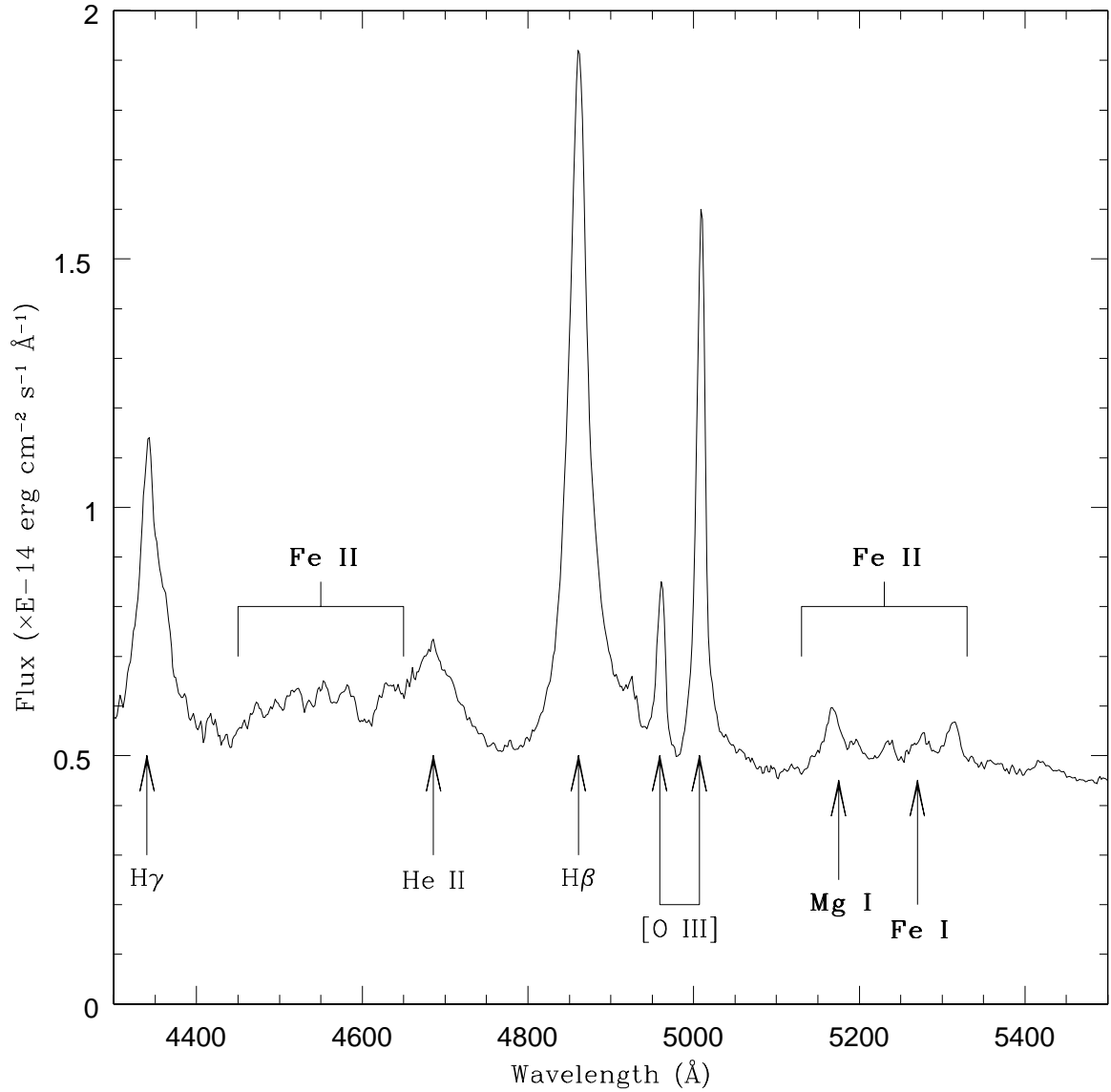}
\caption[Botte.fig5.ps]{Nuclear spectrum of Mrk 335 showing strong \pfeii\, multiplets. 
Arrows indicate emission lines, and the position of stellar absorption 
(Mg I and Fe I).\label{f5}}
\end{figure}

Contrary to what happens for most of S1 galaxies \citep{nel95}, 
measuring $\sigma_*$ in NLS1s with optical spectra is very difficult and 
sometimes even impossible because of the presence of large and bright 
\pfeii\ multiplets, which completely suppress the typically used stellar 
absorption lines, like \eg \fullmgi\ and \fullpfei\ (Fig.~\ref{f5}).
Moreover, NLS1s having bright \pfeii\, show often the 
\pcaii\, triplet ($\lambda \sim 8550$\AA) in emission \citep{pers88}, 
preventing the use of these lines which are generally seen in absorption 
in S1 galaxies.\\
As an alternative \citet{nel96} have shown that the width of the
narrow emission line \fulloiii\, can replace $\sigma_*$, expressed in terms of
$\rm FWHM([O\,{\small~III}]\lambda5007)/2.35$, though the 
correlation between these two quantities is moderately strong with 
considerable scatter.

Assuming that \fulloiii\, profiles are dominated by virial motion in the
bulge potential, these authors investigated some possible secondary influences 
on NLR kinematics. For example, they noticed that Seyfert galaxies 
with high radio luminosity tend to have \fulloiii\, widths broader than what 
expected in case of gravitational motion, because the gas kinematics can 
be influenced by the presence of a radio jet. They also stressed a slight 
tendency for barred and/or disturbed Seyfert galaxies to have broader \fulloiii\,
emission lines. 
Indeed \citet{bahe91} pointed out that the distribution and kinematics 
of near-nuclear gas can be altered during galaxy interactions. 

\begin{figure}
\plotone{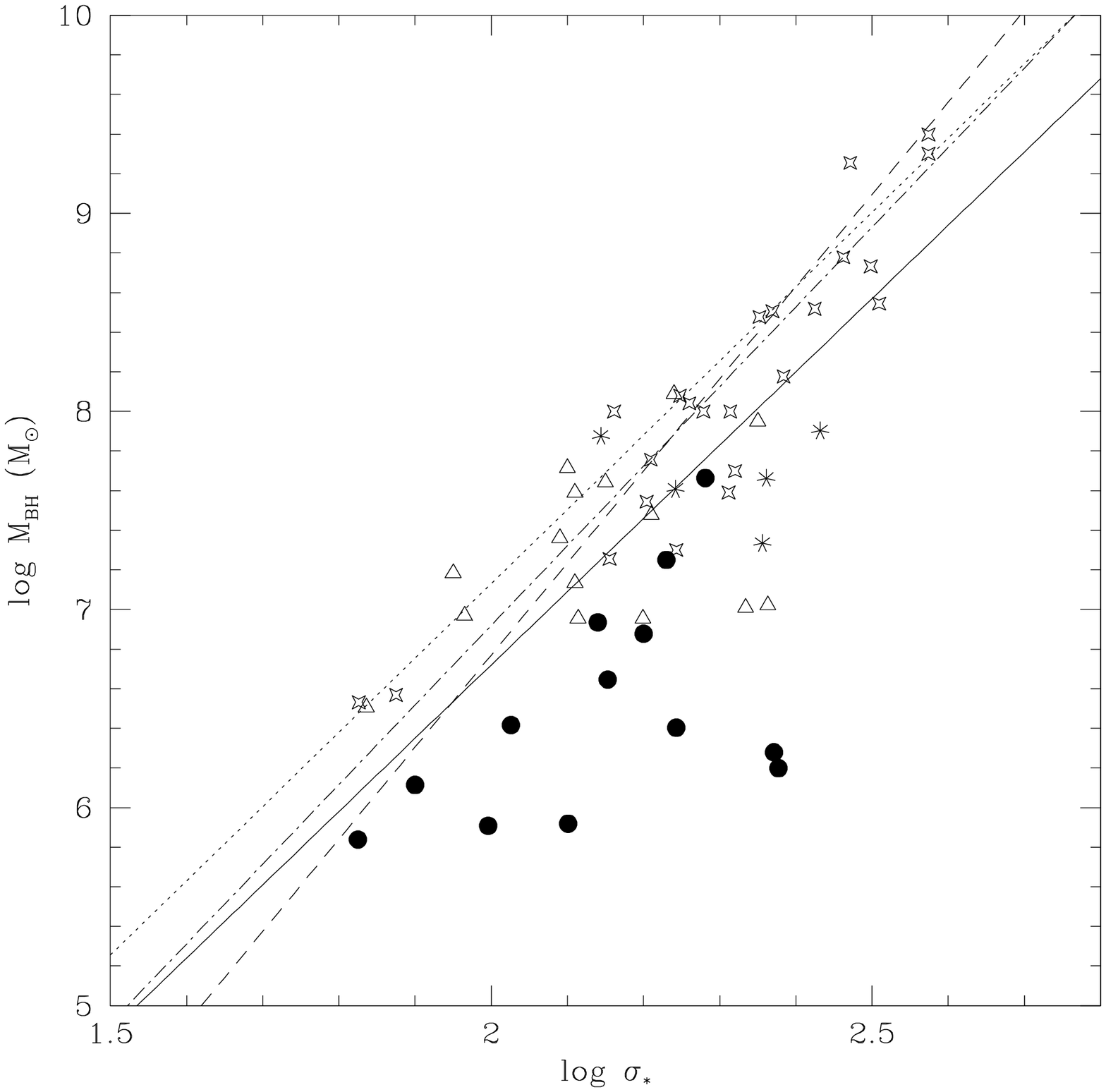}
\caption[Botte.fig6.ps]{${\cal M}_{\rm BH}-\sigma_*$ relation, where $\sigma_*$ is
intended in terms of FWHM(\fulloiii) in all objects for which it was not 
available in literature. Symbols are like in Fig.\ref{f4}, apart from
empty stars which indicate nearby non active galaxies.
The solid line is a least square fit of the total sample of objects:
$\rm log({\cal M}_{BH})=3.70(\pm0.37)~log(\sigma)-0.68(\pm0.80)$. 
The correlation coefficient is R = 0.81.
The relations found by \citet{gebet00a}, \citet{merfer01} and \citet{tre02} are 
also plotted for comparison (dotted line, dashed line and dot-dashed line 
respectively).\label{f6}}
\end{figure}

Since $\sigma_*$ measurments are available in literature only for 
few Seyferts of our sample, we choose to use FWHM(\fulloiii\,) for the others. 
Then for each object we collected the values of their radio luminosity 
($\rm L_{radio}$), taking them from FIRST survey catalog and NED, or , when no 
direct measurements were available, using 
a relationship between the radio luminosity at 1.49\ghz\, and the $\rm L_{FIR}$ 
found by \citet{mapa00} for a sample of Seyfert galaxies: $\rm
log(L_{radio})=0.95(\pm0.06)log(L_{FIR})-12.84(\pm2.07)$ 
\label{radio}. We calculated $\rm L_{FIR}$ using the fluxes at 60 \micron\, and 
100\micron\, extracted from the IRAS Point Source Catalogue and Faint Source 
Catalogue. The total flux $\rm S_{FIR}$ ($40-120$ \micron) was computed by means 
of the relation \citep{he85}: $\rm S_{FIR}=1.26\times\,10^{-14}(2.58S_{60}+S_{100})~~~\rm
W~m^{-2}$, where S$_{60}$ and S$_{100}$ are the flux densities given
in Jansky.  
The so obtained radio luminosities are listed in Tables ~\ref{tab6} 
and \ref{tab7}. The logarithmic values of $\sigma_*$ and ${\cal M}_{BH}$ are 
plotted in Fig.~\ref{f6}, where we have excluded those objects with 
$\rm L_{radio} >  10^{22.5}$ W Hz$^{-1}$, as suggested by \citet{nel96}.
NLS1 galaxies are represented by filled dots and S1s by empty triangles. 
In addition we included quasars from \citet{kaspi00} (asterisks) and nearby non 
active galaxies (empty stars) from \citet{gebet00a}. 

A least square fit of these values (Fig.~\ref{f6}, solid line) gave the 
following relation:

\begin{equation}
\rm log({\cal M}_{BH})=3.70(\pm0.37)~log(\sigma)-0.68(\pm0.80)
\label{sig}
\end{equation}

with a correlation coefficient R$=0.81$.\\ 
This result is in agreement with \citet{ne00} and \citet{walu01}, who found 
${\cal M}_{\rm BH} \propto \sigma_*^{3.70}$, and with \citet{gebet00a}, who 
found ${\cal M}_{\rm BH} \propto \sigma_*^{3.75}$ (Fig.~\ref{f6}, 
dotted line). 
The \citet{merfer01} and \citet{tre02} relations are also plotted in 
Fig.~\ref{f6} for comparison (dashed line and dot-dashed line, respectively).

We notice a larger scatter in our relation with respect to \citet{gebet00a}. 
This is mostly caused by the stellar velocity dispersions of NLS1
galaxies, whose values span a range similar to that of S1s. Contrary to what 
we obtained in \S\ 3.1, this should suggest that NLS1 and S1 galaxies are 
separated when their BH masses are considered, but identical in their bulge
properties. Moreover, all NLS1s remain below the fit 
with some of them closer to the line and, therefore, showing lower 
stellar velocity dispersions corresponding to lower BH masses, while other 
NLS1s have clearly larger $\sigma_*$ than expected. 
This is likely the reason for having a zero point lower than the value obtained 
by \citet{gebet00a}.
A similar result was already obtained by \citet{matet01}, and very recently by 
\citet{bz03b}, who used a large sample of low redshift NLS1s extracted from the 
Sloan Digital Sky Survey, and by \citet{gm04}, who used a sample of NLS1s 
extracted from the ROSAT All-Sky Survey. 
They found that NLS1s mostly deviate from the 
${\cal M}_{\rm BH}-\sigma_*$ relation defined by \citet{tre02}, 
showing $\sigma_*$ values higher than expected.

A possible reason for this scatter could be a spectral resolution 
not sufficient to measure with high precision the low $\sigma_*$ values 
predicted by the fit. This objection is well discussed and rejected by 
\citet{gm04}. Moreover, in our case, those targets observed at higher resolution 
are just NLS1s, while S1s, even if observed at lower resolution, show a similar 
range of $\sigma_*$ values.
Another reason could be the fact that the 
$\rm FWHM([O\,{\small~III}])-\sigma_*$ correlation is not tight and never 
proved to be valid for NLS1s.
Therefore, it is clear that to give a definitive answer, the stellar kinematics 
in NLS1s should be more carefully investigated by means of direct measurments 
of $\sigma_*$. 

Combining (\ref{sig}) with ${\cal M}_{\rm bulge} \propto \sigma_*^{3.3}$  
given by \citet[][and references therein]{wbw00}, who assumed virial 
equilibrium, and
$\rm {\cal M} \propto L^{5/4}$ and $\rm R \propto L^{1/2}$ dependencies, we 
obtain $\rm {\cal M}_{BH}\propto {\cal M}_{bulge}^{1.12\pm0.11}$, which is 
consistent with $\rm {\cal M}_{BH}\propto {\cal M}_{bulge}^{0.85\pm0.09}$ 
given in \S\ 3.1.\\
Moreover starting from equation (\ref{sig}) and converting the BH mass into 
luminosity by means of equation $\rm M_{V}-log({\cal M}_{BH})$ given in 
\S\, 4.1, and the standard relation: $\rm M_{V}~=+4.83-2.5~log(L_V/L_{\odot})$, 
we obtain $\rm L \propto \sigma_*^{3.67}$.
This is perfectly in agreement with the Faber--Jackson relation, 
$\rm L \propto \sigma_*^{n}$, where $\rm n \sim$ 3--4, which is important not 
only in terms of a distance indicator for elliptical galaxies, but also in 
studying the physical properties of bulges, as mentioned by \citet{nel96}.

\section{SUMMARY AND CONCLUSIONS}

In this work we have investigated the nature of NLS1s by following 
an indirect way, that is by using the host galaxy properties to compare black 
hole masses in NLS1s, S1s and also quasars. 

Starting from the assumption that the emission line clouds of BLR are 
gravitationally bound to the BH and in random motion, we have calculated BH 
masses of NLS1s obtaining a typical logarithmic value of 6.65$\pm$0.64
M$_\odot$, which 
is almost one order of magnitude lower than the value obtained for S1s 
(7.37$\pm$0.62 M$_\odot$).
Simultaneously we have confirmed that NLS1s and S1s have quite similar nuclear 
luminosities: (in logarithm), 43.51$\pm$0.65 L$_\odot$ and 43.08$\pm$0.45 
L$_\odot$ respectively. 

The physical properties of the bulges were investigated in NLS1s and S1s by 
means of photometric and spectroscopic data. \\
Published total apparent B magnitudes were corrected for extinction, 
inclination, redshift and AGN contribution, and converted into absolute 
magnitudes. Then, the bulge magnitudes were calculated taking advantage of 
the empirical B/T -- morphology relation given by \citet{side86}, and used 
by several authors.
We plotted these values against BH masses for NLS1s and S1s, adding quasars 
extracted from literature, and we found that NLS1s are mostly confined in the 
lower ranges of the ${\cal M}_{BH}- M_{B}$ plane. This result suggests that 
NLS1s are characterized by less massive BHs hosted in less massive bulges than 
S1s. This is in agreement with previous findings of \citet{walu01}, who 
explored only the ${\cal M}_{\rm BH}-\sigma_*$ relation, and in 
contrast with \citet{matet01} and \citet{bz03}, who claimed for NLS1s with 
lower ${\cal M}_{\rm BH}/{\cal M}_{\rm bulge}$ ratios.
It is not straightforward to justify the different results found by these
authors. After a careful inspection of the effects introduced by each 
correction we applied to the photometric data of our targets, we can observe
that the slightly different way we calculated the correction terms for AGN 
contribution and galaxy inclination seem to cause changes which roughly 
compensate each other. Therefore we guess that the main source of disagreement 
is the morphological classification, which strongly affects the resulting 
bulge magnitudes by quantities in the range 0.25-0.5 dex for uncertainties
discussed in \S\, 3.1.

A fit of NLS1s, S1s and quasar values together shows a strongly correlated 
relation, $\rm M_{B}~=-2.32(\pm0.18)~log({\cal M}_{BH})-3.40(\pm1.36)$,
which leads to $\rm {\cal M}_{BH}\propto {\cal M}_{bulge}^{0.85\pm0.09}$.
Since the slope is close to unity, we do not confirm the results 
by \citet{lao01} and \citet{bz03} who found a significantly non-linear 
BH--bulge correlation.

The velocity dispersion of the bulge stellar component was also explored to 
check whether NLS1s have also $\sigma_*$ values typically lower than those 
measured in S1s.
Since few values of $\sigma_*$ were available in literature, we had to 
calculate the others measuring the \oiii\ emission line widths, and assuming 
that the gas kinematics is dominated by the bulge potential.
As done before, we plotted these values against BH masses, adding quasars and 
also nearby non-active galaxies, and obtaining a good correlation 
${\cal M}_{\rm BH}\propto\sigma_*^{3.70\pm0.37}$.
Contrary to the previous result, NLS1s are not clearly separated from S1s 
in the ${\cal M}_{\rm BH}-\sigma_*$ plane. In particular they span similar 
ranges of $\sigma_*$, suggesting that their bulges are in this respect 
identical. 
Moreover, all NLS1s of our sample fall below the fit, showing $\sigma_*$ values 
higher than expected, and also below the relations found by \citet{gebet00a}, 
\citet{merfer01} and \citet{tre02}, which on the contrary are well in
agreement with our S1s values.
This discrepancy between the two results could be caused by the assumption of 
\oiii\ widths as representative of the stellar kinematics. Indeed the 
\oiii-$\sigma_*$ conversion has a large scatter and is not yet proved to be valid
for NLS1 galaxies. Therefore we stress the necessity to directly measure 
$\sigma_*$, for example observing NLS1s in spectral ranges different form the 
optical one.

Since both relations are based on important assumptions, which can introduce
significant errors, at the moment it is not possible to favor one over the 
other.
The fact that, according to our first result, NLS1 galaxies seem to have less 
massive bulges harboring equally less massive BHs than S1 galaxies, strongly 
indicates that the hypothesis of an inclination effect of a disk-like BLR on 
the narrowness of Balmer emission lines should be rejected, at least for most 
NLS1s. A pure selection effect is not expected to be related to the physical 
properties of the bulges, and therefore NLS1s should be just those S1s with 
intrinsecally smaller bulges to justify the difference we observe between 
NLS1s and S1s. 
On the other hand, our second result confirms the smaller 
$\rm {\cal M}_{BH}/{\cal M}_{bulge}$ ratio in NLS1s, which led \citet{matet01}
and \citet{gm04} to suggest an evolutionary scenario for these AGNs toward a S1 
stage. In particular, NLS1s would be AGNs in a phase when BHs are growing 
independently from their hosting environment.

However, both cases support the general idea of NLS1s in terms of AGNs with 
less massive BHs accreting matter at high rates in order to maintain nuclear 
luminosities comparable to those of S1s, as we said above. 
Moreover, our results are not in conflict with the evolutionary scenario and 
sustain the idea that NLS1s are likely to be young S1s, even in case of a
joined evolution of BH and bulge.
In fact calculations by \citet{bz03} suggest time scales of the order of
some $\sim10^8$ yr for a NLS1 to become a S1. 
This time scale is in agreement with the growth time of a spiral bulge through 
minor merger phenomena. Indeed \citet{wmh96} demostrated 
through N-body simulations that a minor merger makes significant disturbances 
to the morphology of a larger galaxy in less than 1 Gyr of the onset of the 
merger. Moreover \citet{agu01} showed that the accretion of small
satellites is an effective mechanism for the growth of bulges in spiral 
galaxies.

\acknowledgements

We are grateful to the referee for precious comments which improved 
the quality of the paper.
VB is grateful to Y. Lu, S. Kaspi and T. Boller for valuable suggestions. 
SC is grateful to S. Temporin for having supported this work 
with useful discussions.\\ 
This research was partially based on data from the ING Archive.\\
In this work we have used the NASA/IPAC Extragalactic Database (NED) 
which is operated by the Jet Propulsion Laboratory, California Institute of
Technology, under contract with the National Aeronautics and Space 
Administration.\\
This research has made use of the TARTARUS database, which is supported by Jane 
Turner and Kirpal Nandra under NASA grants NAG5-7385 and NAG5-7067.

{}

\clearpage

\clearpage

\begin{deluxetable}{lcccccccc}
\tabletypesize{\scriptsize}
\tablecaption{Narrow-Line Seyfert 1 -- Observation Log. \label{tab1}}
\tablewidth{0pt}
\tablehead{
\colhead{Name} & \colhead{R.A.(J2000)} & \colhead{Dec.(J2000)} & \colhead{Telescope} & 
\colhead{Instrument} & \colhead{$\lambda$ range} & \colhead{Date} & 
\colhead{Exptime} & \colhead{$\Delta \lambda$} \\
& \colhead{(hh mm ss)} & \colhead{(\degr~~~\arcmin~~~\arcsec)} & & & \colhead{(\rm \AA)}
& & \colhead{(sec)} & \colhead{(\rm \AA)}
}
\startdata
Mrk335	& 00 06 19.5 & +20 12 10  & 1.82m & AFOSC & 4000-6500  & 2002 Sep 09 & 1800 & 12 \\
	&            &            &       &       & 6000-8100  & 2002 Sep 09 & 1200 & 9 \\
Mrk957	& 00 41 53.4 & +40 21 18  & INT   & IDS   & 4400-6000  & 1996 Aug 08 & 1200 & 2.5 \\
	&            &            &       &       & 6000-7600  & 1996 Aug 10 & 1200 & 2.5 \\
IZW1	& 00 53 34.9 & +12 41 36  & INT   & IDS   & 4400-6000  & 1996 Aug 08 & 1800 & 2.5 \\
	&            &            &       &       & 5900-7500  & 1996 Aug 07 & 1200 & 2.5 \\
Mrk359	& 01 27 32.5 & +19 10 44  & INT   & IDS   & 4400-6000  & 1996 Aug 08 & 900  & 2.5 \\
	&            &            &       &       & 5900-7500  & 1996 Aug 08 & 900  & 2.5 \\
E0144	& 01 46 44.8 & -00 40 43  & INT   & IDS   & 4400-6000  & 1996 Aug 09 & 1200 & 2.5 \\
	&            &            &       &       & 5900-7500  & 1996 Aug 07 & 1200 & 2.5 \\
Mrk1044 & 02 30 05.4 & -08 59 53  & INT   & IDS   & 4400-6000  & 1996 Aug 08 & 900  & 2.5 \\
        &            &            &       &       & 5900-7500  & 1996 Aug 06 & 900  & 2.5 \\
IR04312 & 04 34 41.5 & +40 14 22  & 1.82m & AFOSC & 4000-6500  & 2002 Sep 10 & 1800 & 12 \\
UGC3478 & 06 32 47.2 & +63 40 25  & 1.82m & AFOSC & 4000-6500  & 2003 Jan 27 & 1800 & 12 \\
        &            &            &       &       & 6000-8100  & 2003 Jan 27 & 1200 & 9 \\
Mrk705  & 09 26 03.3 & +12 44 04  & INT   & IDS   & 4400-6000  & 1995 Jan 17 & 1500 & 3.5 \\
Mrk142	& 10 25 31.3 & +51 40 35  & INT   & IDS   & 4400-6000  & 1995 Jan 17 & 1800 & 4 \\
	&            &            &       &       & 5850-7800  & 1995 Jan 17 & 1800 & 4 \\
IC3599 	& 12 37 41.2 & +26 42 27  & WHT   & ISIS  & 3350-5900  & 1995 Mar 05 & 450  & 9 \\
	&            &            &       &       & 5400-8350  & 1995 Mar 05 & 450  & 2.5 \\
Mrk684	& 14 31 04.8 & +28 17 14  & INT   & IDS   & 4200-7400  & 1996 Aug 11 & 1200 & 6.5 \\
Mrk478	& 14 42 07.4 & +35 26 23  & INT   & IDS   & 4100-7200  & 1996 Aug 10 & 900  & 6.5 \\
Mrk291	& 15 55 07.9 & +19 11 33  & INT   & IDS   & 4400-6000  & 1996 Aug 07 & 900  & 2.5 \\
	&            &            &       &       & 5900-7500  & 1996 Aug 05 & 900  & 2.5 \\
Mrk493	& 15 59 09.6 & +35 01 48  & INT   & IDS   & 4400-6000  & 1996 Aug 06 & 900  & 2.5 \\
	&            &            &       &       & 5900-7500  & 1996 Aug 08 & 900  & 2.5 \\
Hb1557	& 15 59 22.2 & +27 03 39  & INT   & IDS   & 4400-6000  & 1996 Aug 08 & 1200 & 2.5 \\
	&            &            &       &       & 5900-7500  & 1996 Aug 08 & 1200 & 2.5 \\
Kaz163	& 17 46 59.8 & +68 36 39  & INT   & IDS   & 4400-5950  & 1996 Aug 07 & 1200 & 2.5 \\
	&            &            &       &       & 5900-7450  & 1996 Aug 05 & 1200 & 2.5 \\
Mrk507	& 17 48 38.4 & +68 42 16  & INT   & IDS   & 4300-6000  & 1996 Aug 09 & 1200 & 2.5 \\
	&            &            &       &       & 5900-7050  & 1996 Aug 07 & 900  & 2.5 \\
Mrk896	& 20 46 20.9 & -02 48 45  & INT   & IDS   & 4400-5950  & 1996 Aug 08 & 1200 & 2.5 \\
	&            &            &       &       & 5900-7450  & 1996 Aug 06 & 1200 & 2.5 \\
Ark564	& 22 42 39.3 & +29 43 31  & WHT	  & ISIS  & 4700-5550  & 1999 Jun 10 &  600 & 1.5 \\
Ucm2257 & 22 59 32.9 & +24 55 06  & INT   & IDS   & 4250-7450  & 1996 Aug 11 & 1800 & 2.5\\
Mrk1126 & 23 00 47.8 & -12 55 07  & INT   & IDS   & 4400-6000  & 1996 Aug 09 & 1200 & 6.5 \\
        &            &            &       &       & 5900-7500  & 1996 Aug 07 & 1200 & 2.5 \\
\enddata
\end{deluxetable}

\clearpage

\begin{deluxetable}{lcccccccc}
\tabletypesize{\scriptsize}
\tablecaption{Seyfert 1 -- Observation Log. \label{tab2}}
\tablewidth{0pt}
\tablehead{
\colhead{Name} & \colhead{R.A.(J2000)} & \colhead{Dec.(J2000)} & \colhead{Telescope} & 
\colhead{Instrument} & \colhead{$\lambda$ range} & \colhead{Date} & 
\colhead{Exptime} & \colhead{$\Delta \lambda$} \\
& \colhead{(hh mm ss)} & \colhead{(\degr~~~\arcmin~~~\arcsec)} & & & \colhead{(\rm \AA)}
& & \colhead{(sec)} & \colhead{(\rm \AA)}}
\startdata
Mrk1146  & 00 47 19.4 & +14 42 13 & 1.82m & AFOSC & 4000-6500 & 2002 Sep 30 & 1200 & 7.5 \\
         &            &           & 	  &       & 6000-8100 & 2002 Sep 30 & 1200 & 5.5 \\
UGC524   & 00 51 35.0 & +29 24 05 & 1.82m & AFOSC & 4000-6500 & 2003 Jan 26 & 1800 & 12 \\
         &            &           &       &       & 6000-8100 & 2003 Jan 26 & 1200 & 9 \\
Mrk975   & 01 13 51.0 & +13 16 18 & 1.82m & AFOSC & 4000-6500 & 2003 Jan 27 & 1800 & 12 \\
         &            &           &       &       & 6000-8100 & 2003 Jan 27 & 1200 & 9 \\
Mrk358   & 01 26 33.6 & +31 36 59 & 1.82m & AFOSC & 4000-6500 & 2002 Sep 30 & 1800 & 7.5 \\
         &            &           &       &       & 6000-8100 & 2002 Sep 30 & 1200 & 5.5\\
Mrk1040  & 02 28 14.5 & +31 18 42 & 1.82m & AFOSC & 4000-6500 & 2002 Sep 30 & 1200 & 7.5 \\
         &            &           &       &       & 6000-8100 & 2002 Sep 30 & 1200 & 5.5 \\
UGC3142  & 04 43 46.8 & +28 58 19 & 1.82m & AFOSC & 6000-8100 & 2003 Jan 25 & 1800 & 12 \\
         &            &           &       &       & 6000-8100 & 2003 Jan 25 & 1200 & 9 \\
Mrk10	 & 07 47 29.1 & +60 56 01 & INT   & IDS   & 4400-5950 & 1995 Jan 14 & 1500 & 3.5 \\
Mrk382   & 07 55 25.3 & +39 11 10 & 1.82m & AFOSC & 4000-6500 & 2003 Jan 25 & 1800 & 12 \\
         &            &           &       &       & 6000-8100 & 2003 Jan 25 & 1200 & 9 \\
Mrk124	 & 09 48 42.6 & +50 29 31 & INT   & IDS   & 4400-6000 & 1995 Jan 17 & 1500 & 3.5 \\
NGC3080  & 09 59 55.8 & +13 02 38 & INT   & IDS   & 4400-6000 & 1995 Jan 15 & 1800 & 3.5 \\
Mrk40	 & 11 25 36.2 & +54 22 57 & INT   & IDS   & 4400-6000 & 1995 Jan 17 & 1500 & 3.5 \\
Mrk205   & 12 21 44.0 & +75 18 38 & 1.82m & AFOSC & 4000-6500 & 2003 Jan 25 & 1800 & 12 \\
         &            &           &       &       & 6000-8100 & 2003 Jan 25 & 1200 & 9 \\
Ton730 	 & 13 43 56.7 & +25 38 48 & WHT   & ISIS  & 3500-6000 & 1999 Mar 23 & 1200 & 4.0\\
	 &            &           &       &       & 5800-8250 & 1999 Mar 23 & 1200 & 5.5\\
3C382    & 18 35 03.4 & +32 41 47 & WHT   & ISIS  & 3700-5350 & 2000 Jul 06 &  600 & 2.5\\
         &            &           &       &       & 6000-7400 & 2000 Jul 06 &  600 & 4.5 \\
3C390.3  & 18 42 09.0 & +79 46 17 & INT   & IDS   & 4300-5800 & 1995 Jan 14 & 1800 & 3.5 \\
\enddata
\end{deluxetable}

\clearpage

\begin{deluxetable}{lccccccccccc}
\tabletypesize{\scriptsize}
\tablecaption{Photometric data of NLS1 and S1 galaxies. \label{tab3}}
\tablewidth{0pt}
\tablehead{
\colhead{Object} & \colhead{T} & \colhead{$m_{B}$} & \colhead{$\Delta{m_A}$} & 
\colhead{$\Delta{m_i}$} & \colhead{$\Delta{m_g}$} &  \colhead{$\Delta{m_k}$} & 
\colhead{$m_{B}^{\prime}$} & \colhead{$M_B$} & 
\colhead{$\Delta{m_{bulge}}$} & \colhead{$M_{B,bulge}$} 
}
\startdata
\multicolumn{11}{c}{Narrow-Line Seyfert 1}\\
\\
Mrk335  & 0 & 	14.20 &	0.36 &	~0.00     &	-0.15 &	-0.12 & 14.29 &	-20.80 & 0.86 & -19.94 \\
IZw1    & 2 &  	14.47 &	1.56 &	-0.17 &	-0.28 &	-0.18 & 15.41 &	-21.52 & 1.23 &	-20.29 \\
Mrk359  & 3 &   14.52 &	0.67 &	-0.14 &	-0.23 &	-0.04 & 14.78 &	-19.34 & 1.54 &	-17.80 \\
Mrk1044	& 1 &	14.64 &	0.39 &	-0.10 &	-0.15 &	-0.06 & 14.72 &	-19.30 & 1.02 &	-18.28 \\
UGC3478	& 5 &	13.85 &	0.10 &	-0.64 &	-0.39 &	-0.05 & 12.87 &	-20.81 & 2.54 &	-18.27 \\
Mrk142	& 1 &	16.06 &	0.69 &	-0.14 &	-0.07 &	-0.21 & 16.33 &	-19.94 & 1.02 &	-18.92 \\
Mrk684  & 2 &   15.26 &	1.73 &	-0.20 &	-0.09 & -0.14 & 16.56 &	-19.77 & 1.23 &	-18.54 \\
Mrk493  & 4 & 	14.79 &	0.37 &	-0.19 &	-0.11 &	-0.06 & 14.80 &	-20.71 & 1.97 &	-18.74 \\
Mrk896  & 4 &  	15.15 &	0.94 &	-0.37 &	-0.20 &	-0.05 & 15.47 &	-19.66 & 1.97 &	-17.68 \\
Ark564	& 4 &	14.81 &	1.85 &	-0.26 &	-0.26 &	-0.04 & 16.10 &	-18.85 & 1.97 &	-16.88 \\
Mrk590	& 1 &	14.45 &	0.04 &	-0.05 &	-0.16 &	-0.10 & 14.18 &	-20.94 & 1.02 &	-19.92 \\
\\
\multicolumn{11}{c}{Seyfert 1}\\
\\
Mrk1146	& 2 &	15.53 &	0.25 &	-0.23 &	-0.43 &	-0.12 & 15.00 &	-20.97 & 1.23 &	-19.74 \\
UGC524	& 3 &	14.40 &	0.25 &	-0.05 &	-0.27 &	-0.09 & 14.24 &	-21.57 & 1.54 &	-20.02 \\
Mrk975	& 1 &	15.67 &	0.26 &	-0.26 &	-0.11 &	-0.11 & 15.45 &	-21.02 & 1.02 &	-20.01 \\
Mrk358	& 4 &	14.83 &	0.09 &	-0.19 &	-0.21 &	-0.09 & 14.43 &	-21.84 & 1.97 &	-19.87 \\
Mrk1040	& 4 &	13.89 &	0.20 &	-1.32 &	-0.41 &	-0.03 & 12.33 &	-21.77 & 1.97 &	-19.80 \\
Mrk10	& 4 &	14.53 & 0.18 &	-0.63 &	-0.20 &	-0.06 & 13.82 &	-21.52 & 1.97 &	-19.55 \\
Mrk382	& 3 &	15.30 &	0.24 &	-0.06 &	-0.21 &	-0.06 & 15.18 &	-20.44 & 1.97 &	-18.47 \\
Mrk124	& 3 &	15.94 &	0.53 &	-0.24 &	-0.06 &	-0.13 & 16.04 &	-20.73 & 1.54 &	-19.19 \\
NGC3080	& 1 &	15.70 &	0.40 &	-0.03 &	-0.13 &	-0.13 & 15.82 &	-19.95 & 1.02 &	-18.93 \\
Mrk205	& 1 &	15.23 &	0.20 &	-0.25 &	-0.18 &	-0.26 & 14.74 &	-22.51 & 1.02 &	-21.49 \\
Mrk817	& 1 &	14.33 &	0.46 &	 ~0.00    &	-0.03 &	-0.12 & 14.64 &	-20.85 & 1.02 &	-19.83 \\
NGC3516	& 0 &	12.59 &	0.15 &	-0.08 &	-0.18 &	-0.04 & 12.44 &	-20.29 & 0.86 &	-19.43 \\
\enddata
\tablecomments{Column:
(1) -- object name; 
(2) -- morphological type; 
(3) -- total apparent B magnitude; 
(4,5,6,7) -- magnitude corrections for nuclear nonstellar continuum and 
emission lines ($\Delta{m_A}$), inclination ($\Delta{m_i}$), Galactic 
absorption ($\Delta{m_g}$), and redshift ($\Delta{m_k}$); 
(8) -- total corrected apparent B magnitude; 
(9) -- total corrected absolute B magnitude; 
(10) -- correction to obtain bulge magnitude; 
(11) -- bulge absolute B magnitude. 
}
\end{deluxetable}

\clearpage

\begin{deluxetable}{lcccccccc}
\tabletypesize{\scriptsize}
\tablecaption{Quasar sample. \label{tab4}}
\tablehead{
\colhead{Name} & \colhead{R.A.} & \colhead{Dec.} & \colhead{z} & 
\colhead{$f_{\lambda}(5100$\AA)} & \colhead{$M_{B}$} & 
\colhead{$log~L_{1415}$} & \colhead{${\cal M}_{BH}$} & \colhead{$log~\sigma_*$} \\
  & \colhead{(hh mm ss)} & \colhead{(\degr~~~\arcmin~~~\arcsec)} & & \colhead{(\ergcmsa)} & & 
  \colhead{(\Whz)} & \colhead{($10^7M_{\odot}$)} & \colhead{(\kms)}
}
\startdata
PG0026\ldots & 00 29 13.6 & +13 16 03 & 0.142 &  $2.7\times10^{-15}$ & -22.82 & 23.71 & $ 5.4^{+1.0}_{-1.1}$ & \nodata \\
PG0052\ldots & 00 54 52.1 & +25 25 38 & 0.155 &  $2.1\times10^{-15}$ & -23.68 & 22.95 & $22.0^{+6.3}_{-5.3}$ & \nodata \\
PG0804\ldots & 08 10 58.6 & +76 02 42 & 0.100 &  $5.5\times10^{-15}$ & -22.86 & 22.64 & $18.9^{+1.9}_{-1.7}$ & \nodata \\
PG0844\ldots & 08 47 42.4 & +34 45 04 & 0.064 &  $3.7\times10^{-15}$ & -21.01 & 21.79 & $2.16^{+0.90}_{-0.83}$ & 2.356\\
PG0953\ldots & 09 56 52.4 & +41 15 22 & 0.239 &  $1.6\times10^{-15}$ & -24.03 & 23.77 & $18.4^{+2.8}_{-3.4}$  & \nodata  \\
PG1211\ldots & 12 14 17.7 & +14 03 13 & 0.085 &  $5.7\times10^{-15}$ & -21.12 & 22.47 & $4.05^{+0.96}_{-1.21}$ & 2.242\\
PG1226\ldots & 12 29 06.7 & +02 03 09 & 0.158 & $21.3\times10^{-15}$ & -24.11 & 27.67 & $55.1^{+8.9}_{-7.9}$ & \nodata  \\
PG1229\ldots & 12 32 03.6 & +20 09 29 & 0.064 &  $2.1\times10^{-15}$ & -22.18 & 22.12 & $7.5^{+3.6}_{-3.5}$ & 2.144\\
PG1307\ldots & 13 09 47.0 & +08 19 49 & 0.155 &  $1.8\times10^{-15}$ & -22.78 & 22.63 & $28^{+11}_{-18}$   & \nodata \\
PG1351\ldots & 13 53 15.8 & +63 45 45 & 0.087 &  $5.1\times10^{-15}$ & -22.19 & 22.29 & $4.6^{+3.2}_{-1.9}$ & 2.361\\
PG1411\ldots & 14 13 48.3 & +44 00 14 & 0.089 &  $3.7\times10^{-15}$ & -21.96 & 22.38 & $8.0^{+3.0}_{-2.9}$ & 2.432 \\
PG1613\ldots & 16 13 57.2 & +65 43 10 & 0.129 &  $3.5\times10^{-15}$ & -22.86 & 23.41 & $24.1^{+12.5}_{-8.9}$ & \nodata \\
PG1617\ldots & 16 20 11.3 & +17 24 28 & 0.114 &  $1.4\times10^{-15}$ & -22.47 & 22.85 & $27.3^{+8.3}_{-9.7}$  & \nodata \\
PG2130\ldots & 21 32 27.8 & +10 08 19 & 0.061 &  $4.8\times10^{-15}$ & -21.48 & 22.56 & $14.4^{+5.1}_{-1.7}$  & \nodata \\
\enddata
\tablecomments{Column:
(1) -- object name;
(2,3) -- right ascension and declination (J2000);
(4) -- redshift;
(5) -- continumm flux at $\sim5100$\AA;
(6) -- absolute B magnitude;
(7) -- logarithmic radio luminosity at 1415 \mhz\ \citep{ne00};
(8) -- estimated black hole mass in units of $(10^7M_{\odot})$;
(9) -- logarithmic stellar velocity dispersion obtained from FWHM([O III])
values by \citet{ne00}.
}
\end{deluxetable}

\clearpage

\begin{deluxetable}{llccc} 
\tabletypesize{\scriptsize}
\tablecaption{${\cal M}_{BH}\propto{\cal M}_{bulge}^{\alpha}$ for 
different authors. \label{tab5}}
\tablewidth{0pt}
\tablehead{
\colhead{Reference} & \colhead{Sample} & \colhead{N} & \colhead{$\alpha$} & 
\colhead{R}
}
\startdata
This work & NLS1s, S1s  and quasars & 37 & $0.85\pm0.09$ & 0.91 \\
Bian \& Zhao (2003)  & NL AGNs & 22 & $1.61\pm0.59$ & 0.74\\
Wandel (2002) & AGNs & 47 & $0.74\pm 0.11$ & 0.67\\
McLure \& Dunlop (2002) & AGNs & 72 & $0.88\pm 0.06$ & 0.77\\
Laor (2001) & quasars \& Seyfert & 24 & $1.36\pm 0.21$ & 0.80 \\
Laor (2001) & AGNs + Quiescent Galaxies& 40 & $1.54\pm 0.15$ & 0.80 \\
\enddata
\tablecomments{Column:
(1) -- bibliographic reference;
(2) -- sample;
(3) -- number of objects;
(4) -- exponent of ${\cal M}_{BH}\propto{\cal M}_{bulge}^{\alpha}$ relation;
(5) -- correlation coefficient.
}
\end{deluxetable}

\clearpage

\begin{deluxetable}{lrrrcrcl}
\tabletypesize{\scriptsize}
\tablecaption{Narrow-Line Seyfert 1 -- Black hole masses and stellar velocity 
dispersions. \label{tab6}}
\tablewidth{0pt}
\tablehead{
\colhead{NLS1} & \colhead{z} & \colhead{$\lambda L_{\lambda}(\rm 5100\AA)$} & 
\colhead{$R_{BLR}$} & \colhead{$FWHM(H\beta)$}  &   \colhead{$M_{BH}$} & 
\colhead{$log~L_{radio}$} & \colhead{$log~\sigma_*$} \\
  &  & \colhead{(\ergsec)} & \colhead{(lt days)} & \colhead{(\kms)} & 
  \colhead{($10^6M_{\odot}$)} & \colhead{(\Whz)} & \colhead{(\kms)} 
}
\startdata
Mrk335	& 0.0260 & 3.1 $10^{43}$  & 14.7$\pm$2.4 &  2007.7   &  8.61$\pm$0.84 & 21.99\tablenotemark{a} & 2.140 \\
IZW1	& 0.0606 & 29.2 $10^{43}$ & 69.9$\pm$6.6 &  1439.2   & 17.00$\pm$5.06 & 22.79\tablenotemark{a} & 2.769 \\
Mrk359	& 0.0167 & 1.1 $10^{43}$  &  7.1$\pm$0.9 &   816.2   &  0.69$\pm$0.09 & 21.45\tablenotemark{a}& 1.825 \\
E0144	& 0.0827 & 2.5 $10^{43}$  & 12.6$\pm$1.3 &  1566.2   &  4.50$\pm$0.47 & 22.75\tablenotemark{b}& 2.132 \\
Mrk1044	& 0.0159 & 0.6 $10^{43}$  &  4.6$\pm$0.7 &  1684.6   &  1.90$\pm$0.29 & 21.09\tablenotemark{b}& 2.371  \\
IR04312	& 0.0201 & 3.25 $10^{43}$ & 15.0$\pm$1.4 &  1373.7   &  0.83$\pm$0.08 & 22.14\tablenotemark{a} & 2.101 \\
UGC3478 & 0.0131 & 0.3 $10^{43}$  &  2.6$\pm$0.4 &  1460.8   &  0.81$\pm$0.10 & 21.62\tablenotemark{a} & 1.996 \\
Mrk705  & 0.0287 & 29.2 $10^{43}$ & 68.3$\pm$6.3 &  2153.5   & 46.13$\pm$4.31 & 22.13\tablenotemark{b} & 2.281 \\
Mrk142	& 0.0448 & 3.3  $10^{43}$ & 15.2$\pm$1.4 &  1876.3   &  7.54$\pm$0.72 & 21.36\tablenotemark{b} & 2.200 \\
IC3599	& 0.0224 & 0.1  $10^{43}$ &  1.7$\pm$0.3 &   692.9   &  0.13$\pm$0.02 & \nodata & 2.055 \\
Mrk684	& 0.0463 & 8.7 $10^{43}$  &  9.8$\pm$1.9 &  1805.6   & 14.17$\pm$0.91 & 22.52\tablenotemark{c} & 2.669 \\
Mrk478	& 0.0774 & 28.9 $10^{43}$ & 69.4$\pm$6.5 &  1979.2   & 30.60$\pm$3.34 & 22.58\tablenotemark{b} & 2.786 \\
Mrk493	& 0.0316 & 2.1  $10^{43}$ & 11.1$\pm$1.2 &   989.9   &  1.58$\pm$0.02 & 21.81\tablenotemark{b} & 2.377 \\
Kaz163	& 0.0637 & 3.3 $10^{43}$  & 15.3$\pm$1.5 &  1844.8   &  8.76$\pm$0.83 & 22.80\tablenotemark{c} & 2.332 \\
Mrk507	& 0.0553 & 3.2 $10^{43}$  & 15.0$\pm$1.4 &  1876.3   &  9.44$\pm$1.04 & 22.52\tablenotemark{c} & 2.519 \\
Mrk896	& 0.0265 & 2.0 $10^{43}$  & 10.6$\pm$1.1 &  1673.4   &  4.43$\pm$0.48 & 21.47\tablenotemark{c} & 2.153 \\
Ark564	& 0.0244 & 4.2 $10^{43}$  & 17.9$\pm$1.6 &   971.6   &  2.61$\pm$0.26 & 22.00\tablenotemark{c} & 2.026 \\
Ucm2257 & 0.0336 & 1.6 $10^{43}$  &  9.2$\pm$1.1 &  1375.3   &  2.53$\pm$0.30 & 22.47\tablenotemark{c} & 2.243 \\
NGC4051	& 0.0023 & \nodata         &  \nodata      & \nodata     &  1.30$\pm$0.10\tablenotemark{d} & 20.29\tablenotemark{b}  & 1.90\tablenotemark{e} \\
Mrk590	& 0.0264 & \nodata         &  \nodata      & \nodata     &  17.8$\pm$0.38\tablenotemark{d} & 22.12\tablenotemark{b}  & 2.23\tablenotemark{e} \\
\enddata
\tablecomments{Column:
(1) -- object name;
(2) --  redshift;
(3) -- continuum luminosity at $\rm 5100\AA$;
(4) -- the estimated size of broad line region in light-days;
(5) -- $H\beta$ line width in \kms;
(6) -- the estimated black hole mass;
(7) -- the logarithmic value of stellar velocity dispersion measured from FWHM([O III]$\lambda$5007).
} 
\tablenotetext{a}{$L_{radio}$ taken from NED}
\tablenotetext{b}{$L_{radio}$ taken from FIRST catalog}
\tablenotetext{c}{$L_{radio}$ derived from $L_{FIR}$}
\tablenotetext{d}{Black hole masses taken from \citet{Wuhan01}} 
\tablenotetext{e}{Stellar velocity dispersions taken from \citet{wan02}}
\end{deluxetable}

\clearpage

\begin{deluxetable}{lrrrcrcl}
\tabletypesize{\scriptsize}
\tablecaption{Seyfert 1 -- Black hole masses and stellar velocity 
dispersions. \label{tab7}}
\tablewidth{0pt}
\tablehead{
\colhead{S1} &  \colhead{z} & \colhead{$\lambda L_{\lambda}(\rm 5100\AA)$} & 
\colhead{$R_{BLR}$} & \colhead{$FWHM(H\beta)$} & \colhead{$M_{BH}$} & 
\colhead{$log~L_{radio}$} & \colhead{$log~\sigma_*$} \\
 &  &  \colhead{(\ergsec)}  &  \colhead{(lt days)}  & \colhead{(\kms)} & 
 \colhead{($10^7M_{\odot}$)} &  \colhead{(\Whz)} & \colhead{(\kms)} 
}
\startdata
Mrk1146 &0.0391	&  1.1 $10^{43}$  & 7.2$\pm$0.9  &   3141.6  &   1.05$\pm$1.35  & 21.70\tablenotemark{a}  & 2.363 \\
UGC524	&0.0366	&  1.6 $10^{43}$  & 9.3$\pm$1.1  &   4160.6  &   2.36$\pm$0.34  & 22.62\tablenotemark{a}  & 2.350 \\
Mrk975	&0.0492	&  1.4 $10^{43}$  & 8.3$\pm$1.0  &   3775.4  &   1.71$\pm$0.24  & 22.75\tablenotemark{a}  & 2.577 \\
Mrk358  &0.0452	&  0.7 $10^{43}$  & 5.2$\pm$0.7  &   2234.8  &   0.32$\pm$0.05  & 22.32\tablenotemark{c}  & 1.836 \\
Mrk1040 &0.0164	&  0.8 $10^{43}$  & 5.7$\pm$0.8  &   4220.1  &   1.36$\pm$0.20  & 21.90\tablenotemark{a}  & 2.110 \\
UGC3142	&0.0218	&  1.0 $10^{43}$  & 6.6$\pm$1.7  &  10488.5  &   8.91$\pm$0.22  & 22.36\tablenotemark{a}  & 2.350 \\
Mrk10	&0.0293	&  1.2 $10^{43}$  & 7.5$\pm$0.9  &   3046.9  &   0.90$\pm$0.12  & 22.27\tablenotemark{c}  & 2.114 \\
Mrk382	&0.0332	&  1.2 $10^{43}$  & 7.4$\pm$0.9  &   2903.3  &   0.90$\pm$0.07  & 21.86\tablenotemark{c}  & 2.199 \\
Mrk124	&0.0564	&  3.0 $10^{43}$  & 14.1$\pm$1.3 &   1982.3  &   0.73$\pm$0.02  & 22.55\tablenotemark{b}  & 2.411\\
NGC3080	&0.0355	&  1.1 $10^{43}$  & 6.9$\pm$0.9  &   3172.4  &   1.02$\pm$1.03  & 22.06\tablenotemark{c}  & 2.334 \\
Mrk40	&0.0206	&  0.3 $10^{43}$  & 3.5$\pm$0.8  &   4042.3  &   0.93$\pm$0.13  & 21.10\tablenotemark{b}  & 1.965 \\	
Mrk205	&0.0703	&  4.9 $10^{43}$  & 19.9$\pm$1.7 &   5082.9  &   7.48$\pm$0.31  & 22.67\tablenotemark{c}  & 2.275 \\
Ton730	&0.0853	&  2.4 $10^{43}$  & 11.8$\pm$1.6 &   3761.6  &   2.43$\pm$0.17  & 22.78\tablenotemark{c}  & 2.233 \\
3C390.3	&0.0559	&  5.5 $10^{43}$  & 21.8$\pm$1.6 &   5284.7  &   22.3$\pm$2.4   & 25.82\tablenotemark{a}  & 2.336 \\
3C382   &0.0559	& 23.6 $10^{43}$  & 60.3$\pm$5.3 &  15991.4  & 224.60$\pm$16.31 & 25.49\tablenotemark{a}  & 2.616 \\
Mrk79   &0.0222 &  \nodata         & \nodata       & \nodata     &   5.20$\pm$2.40\tablenotemark{d}  & 22.15\tablenotemark{b}  & 2.10\tablenotemark{e}  \\
3C120	&0.0330	&  \nodata   	   & \nodata       & \nodata     &   3.00$\pm$1.30\tablenotemark{d}  & \nodata               & 2.21\tablenotemark{e}  \\	
Mrk817	&0.0314	&  \nodata         & \nodata 	   & \nodata     &   4.40$\pm$1.20\tablenotemark{d}  & 22.24\tablenotemark{b}  & 2.15\tablenotemark{e} \\
NGC3227	&0.0038	& \nodata          & \nodata       & \nodata     &   3.90$\pm$3.00\tablenotemark{d}  & 21.36\tablenotemark{b}  & 2.11\tablenotemark{e} \\
NGC3516	&0.0088	& \nodata 	   & \nodata       & \nodata     &   2.30$\pm$0.90\tablenotemark{d}  & \nodata               & 2.09\tablenotemark{e} \\
NGC5548	&0.0171 & \nodata 	   & \nodata       & \nodata     &  12.30$\pm$1.60\tablenotemark{d}  & 22.14\tablenotemark{b}  & 2.24\tablenotemark{e}  \\
NGC4151	&0.0033	& \nodata          & \nodata       & \nodata     &   1.53$\pm$0.93\tablenotemark{d}  & 21.84\tablenotemark{b}  & 1.95\tablenotemark{e}  \\
\enddata
\tablecomments{As in Table \ref{tab6}}
\end{deluxetable}

\end{document}